# Approximate String Matching for DNS Anomaly Detection

Roni Mateless and Michael Segal
Ben-Gurion University of the Negev, Israel

mateless@post.bgu.ac.il, segal@bgu.ac.il

**Abstract.** In this paper we propose a novel approach to identify anomalies in DNS traffic. The traffic time-points data is transformed to a string, which is used by new fast approximate string matching algorithm to detect anomalies. Our approach is generic in its nature and allows fast adaptation to different types of traffic. We evaluate the approach on a large public dataset of DNS traffic based on 10 days, discovering more than order of magnitude DNS attacks in comparison to auto-regression as a baseline. Moreover, the additional comparison has been made including other common regressors such as Linear Regression, Lasso, Random Forest and KNN, all of them showing the superiority of our approach.

## 1 Introduction

Domain Name System (DNS) is the white pages of the Internet allowing to map hostnames/domains to IP addresses and it facilitates the communication of devices over the Internet. DNS is based on Standards (RFCs 1034, 1035-> STD13, Updated by a number of RFCs). It's a distributed and hierarchical database. Since DNS is an open ASCII protocol with no encryption, it is quite vulnerable for many known security holes. It uses a very rudimentary authentication mechanism which is based only on the SIP, port and transaction ID and, therefore, naively trusts source of information. Caching allows to bypass authoritative records and to store unreliable information in many locations in the internet.

Recently, it was reported that FireEye's Mandiant Incident Response and Intelligence teams have identified a wave of DNS hijacking that has affected dozens of domains belonging to government, telecommunications and internet infrastructure entities across the Middle East, North Africa, Europe and North America [33]. Additionally, the Internet Corporation for Assigned Names and Numbers (ICANN), which supervises the DNS reported on February 2019 that the DNS servers for a variety of prominent targets across the world have been subject to a rush of attacks known as DNS hijacking attacks [34].

Several known malicious actions detectable by analyzing DNS data are known to the community: (D)DoS, Cache poisoning, Tunneling, Fast flux, Zone transfer hijacking, Dynamic update corruption and few others. To address the serious concerns related to the DNS attacks, we introduce unsupervised learning method based on string matching algorithm. We chose to deal with the unsupervised method due to the existence of large amount of unlabeled traffic data. Our approach is general and fast (it runs in linear time), easy to extend to other protocols and easy to maintain because there is no model to train.

In the past, to identify and prevent above-mentioned attacks, a number of statistical indicators for DNS traffic were considered: increase in number of DNS packets, decrease in cache hit ratio, increase in average DNS queries of individual source IP addresses, increase in num-ber of recursive queries, increase in number of source IP addresses within a limited time slot and decrease in ratio of resolved queries. We went one step further by considering several statistical indicators in relation with potentially periodic data traffic. Our method is quite generic and can be applied to other types of traffic as well. In particular, our scheme performs well in identifying all the events related to DDoS and compromising servers attacks. The contributions of this work are the following:

1. A new fast, generic approximate string matching algorithm for DNS anomaly detection.
2. Simulative comparison between our technique and auto-regressive unsupervised scheme and other common regressors such as Linear Regression, Lasso, Random Forest and KNN, using DARPA 2009 dataset from IMPACT Cyber database.
3. Simultaneous use of scale dependent and invariant similarity measures (mean squared error and cosine similarity) to predict anomaly in the DNS data traffic.

The general idea of our solution is to identify similar periodic instances of investigated statistical indicators on the fly and to make a predictive suggestion based on historical data. The prediction is made by our extended approximate string matching algorithm based on KMP algorithm [32]. We can also control the length of historical data that we use during the algorithm. As said above, our solution is very fast in terms of time and memory complexity, and can be applied to different types of traffic. Moreover, as it is shown by experimental results (Section 5), it greatly outperforms (in terms of precision) auto-regression model with many variables, Linear Regression, Lasso, Random Forest, KNN and quite speedy.

This paper is organized as follows. We start with related work on DNS anomalies, then we present our approach and the solution with explanations. Next, we evaluate the proposed solution analytically (Section 4) and experimentally (Section 5) comparing it with auto-regression approach and other common regressors such as Linear Regression, Lasso, Random Forest and KNN. Finally, we conclude the paper and suggest further research.

## 2 Related work

A number of research works have been dealing with identifying of DNS anomalies. As mentioned above, popular technique used by cyber-criminals to hide their critical systems is fast-flux. The ICANN Security and Stability Advisory Committee [1] released a paper giving a clear explanation of the technique .Jose Nazario and Thorsten Holz [2] did some interesting measurements on known fast-flux domains. Villamarn-Salomn and Brustoloni [3] focused their detection on abnormally high or temporally concentrated query rates of dynamic DNS queries. The research by Choi et al. [4] created an algorithm that checks multiple botnet characteristics. Born and Gustafson [5] researched a method for detecting covert channels in DNS using character frequency analysis. Karasaridis [6] used the approach of histograms' calculations of request/response packet sizes using the fact that tracking and detecting the changes in the frequencies of non-conforming packets sizes lead to possible identification of DNS anomaly. Yuchi et al. [7] investigated DNS anomalies in the context of Heap’s law stating that a corpus of text containing $N$ words typically contains on the order of $cN^{\beta}$ distinct words, for constant $c$ and $0<\beta<1$. Cermák et al. [8] identified that only four DNS packet fields are useful for most of the DNS traffic analzing methods: queried domain name, queried record type, response code and IP address returned. Yarochkin et al. [9] presented an open source DNSPACKETLIZER tool to analyze DNS traffic in-real time. The analyzer calculates a Statistical Feature String for every event and stores it along with the query domain name and IP addresses (if such are known). The papers [12,13] investigated some statistical features of the domain name in the context of DNS traffic. The research described in [10] aimed detect the botnet traffic by inspecting the following parameters: Time-Based Features (Access ratio), DNS Answer-Based Features (Number of distinct IP addresses), TTL Value-Based Features, Domain Name-Based Features (% of numerical in domain name). Satam et al. [11] used a dnsgram which is a data structure the captures important properties of consecutive DNS queries and replies. In [14] the authors presented an approach in which the flow of DNS traffic between the source and the destination DNS server is used to detect attacks. For taking care of Feature-Based Detection, variations of entropy based learning mechanisms were developed [15-18]. Based on the definition of context, there is a cause-effect relation among the features that characterize the context *C* and the corresponding consequences.

We note that some past attempts were made in order to bring unsupervised machine learning mechanisms to find DNS related anomalies. Raghuram et al. [19] proposed a method for detecting anomalous domain names, with focus on algorithmically generated domain names which are frequently associated with malicious activities. They [19] used the well-known maximum likelihood estimation (MLE) framework wherein the parameters of a probability model are found by maximizing the likelihood of a training data set under that model. Kirchler et al. [20] presented modified *k*-means algorithm and evaluated it on a realistic dataset that contains the DNS queries. Chatzis and Popescu-Zeletin [22] study the effect email worms have on the flow-level characteristics of DNS query streams a user machine generates using similarity search over time series analysis. The authors in [21] show that they can correctly classify DNS traffic using clustering (k-means) analysis.

# 3 Our Approach

Our unsupervised approach is based on approximate string matching strategy as explained in the following subsections.

## 3.1 General approach

We consider the historical traffic feature (for example, the total number of DNS packets measured per each minute) as the text *T*, the last load traffic measurements as the pattern *P* and the currently measured load as *E*. The idea is to find a set *S* of the starting indices of non-overlapping approximate appearances of *P* in *T* where the absolute difference between any two elements in *S* as at least $|P|$. Then we make a prediction *Pred* based on *S* and compare it to *E*. What does it mean "approximate appearance" of *P* in *T*? We use it to identify a possible periodicity in the history. In particular, we set up different values, $\alpha$ and $\beta$ to control the approximate appearance of *P* in *T*. The parameter $\alpha$ controls the possible difference between each measurement in *P* versus corresponding measurement in *T*, while the parameter $\beta$ gives an upper bound on the total difference between all the measurements in *P* versus corresponding measurements in *T*. After the sequence *S* is found, we make a prediction *Pred* based on the average value of the measurements that are located immediately after the indices in *S*. (We also consider the boundary cases as well, for example the cold-start case when the sequence *S* could be empty as well). Visually, our solution works as shown in Figure 1.

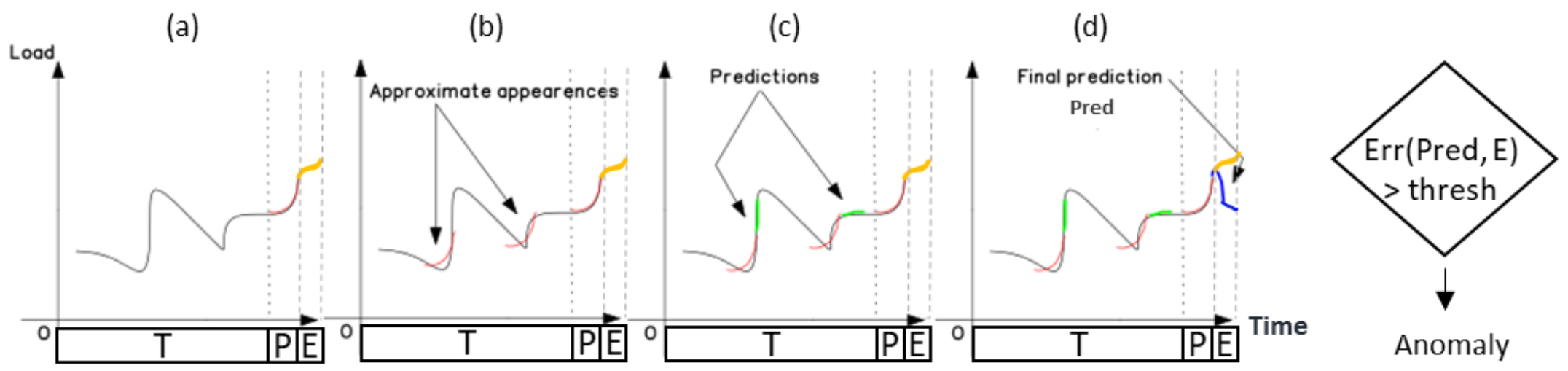

**Fig. 1.** (a) shows the traffic load as the function of time, where the red part is the last load pattern *P* and the orange part is currently measured load *E* ; (b) the algorithm looks for the approximate appearances of *P* in *T*; (c) the predictions (in green) are made based on appearances we found in (b) stage; (d) the prediction *Pred* (in blue) for the currently measured traffic load *E* is made; if no correlation between *Pred* and *E* exists, the anomaly is reported.

In order to understand whether the anomaly happens in the current time, we compare the current measurements *E* with *Pred* as explained in the above paragraph. The comparison is made using the cosine similarity measure and the mean squared error.

### 3.2 The Approximate KMP algorithm

Below we present our matching algorithm and elaborate on the details of proposed solution. We must point out that approximate string matching strategy is based on Knuth-Morris-Pratt fast string matching algorithm [32], but can work, in fact, with any standard string pattern matching solution. We will also show how to use the algorithm in an incremental fashion in order to speed up the running time.

The modified function π is defined as follows:

```
π(P, α):
    ret = [0]
    for i in range(1, len(P)):
      j = ret[i - 1]
      while j > 0 and abs(P[j] - P[i]) > α:
        j = ret[j - 1]
      ret.append(j + 1 if abs(P[j] - P[i]) <= α else j)
return ret
```

The **total_error** function computes the total difference between the pattern *P* and text *T* at given offset:

```
total_error(P, T, offs):
    total = 0
    for i in range(len(P)):
      total += abs(P[i] - T[offs + i])
return total
```

Finally, the approximate searching solution based on Knuth-Morris-Pratt algorithm and modified π function was designed. The returned array "ret" of indices contains the starting indices in text *T* where the approximate matches of *P* are found. The presented algorithm has, obviously, linear (in terms of *T* and *P* sizes) running time.

```
search(T, P, α, β):
    π, ret, j = π(P, α), [], 0
    for i in range(len(T)):
      while j > 0 and abs(T[i] - P[j]) > α:
        j = π[j - 1]
      if abs(T[i] - P[j]) <= α: j += 1
      if j == len(P):
        temp_j = j
        j = π[j - 1]
        if total_error(P, T, i-len(P)+1) <= β:
          ret.append(i - (temp_j - 1))
          j=0
return ret
```

### 3.3 Identifying anomaly

We have used mean squared error and cosine similarity measures (plus additional parameters) in order to make the decision about a possible anomaly in the current data traffic. Assuming *Pred* is a vector of *k* predicted values and *E* is a vector of *k* observed values, the mean squared error is defined as:

$$MSE = \frac{1}{k}\sum_{i=1}^{k}(Pred_i - E_i)^2 \qquad (1)$$

Mean squared error represents the difference between the actual measurements and the measurement values predicted by the model. The choice of cosine similarity measure has been done due to the fact that cosine similarity is a scale invariant similarity measure. It is defined as:

$$\cos(Pred, E) = \frac{\sum(Pred_i \cdot E_i)}{\sqrt{Pred_i^2} \cdot \sqrt{E_i^2}} \qquad (2)$$

However, the cosine similarity measure assumes that both vectors *Pred* and *E* are non-zero vectors which can be problematic in the context of measuring traffic, since it is possible sometimes to obtain zero values. In order to overcome this difficulty, we consider only the case when the vector *Pred* is non-zero vector. This is because no anomaly is expected in data when vector *E* contains only zero values. Also, for the case when the *Pred* vector does not exist (e.g. a cold-start case when no approximate matching has been found), we compare the currently observed measurements vector *E* with the pattern *P*. We note that these measurements of vector *E* come exactly after the occurrence of pattern *P* at the end of data *T*. In case, where vector *E* values are significantly (order of magnitude) larger than vector *P* values, it may indicate a possible anomaly in data.

## 4 Analysis

First, we show the evaluation of our proposed technique to ensure that it does not suffer from cold-start problem. Next, we explain how to make our algorithm incrementally faster without the need of actual recomputing the data.

### 4.1 Cold-start evaluation

We are interested to analyze whether we can experience the problem of cold-start when applying our approximate matching solution. We have an alphabet of $10^d$ letters (every value in pattern *P* is considered to be a letter), and since the length of *P* is *k*, the number of possible matching patterns of *P* in *T* without any error is $10^{dk}$. If we ignore the parameter *β*, the number of choices without *β* is $(2\alpha + 1)^k$ and the total number of choices with parameter *β* is:

$$\beta \geq \binom{k}{\left\lfloor \frac{\beta}{\alpha} \right\rfloor} \cdot (2\alpha + 1)^{\left\lfloor \frac{\beta}{\alpha} \right\rfloor} \cdot \max(1, \beta \bmod \alpha) \quad (3)$$

In fact, if we assume that $d = 3, k = 5$, the number of possible patterns will be $10^{15}$. When we take *α* and *β* to characterize 10% and 25% of shift, respectively, of the maximum measurement value that can be obtained (in other words $\alpha = 100, \beta = 250$), the number of choices for the particular approximate search of pattern *P* inside of historical data *T* is equal to at least $\binom{5}{2} \cdot (201)^2 \cdot 50 \approx 10 \cdot 4 \cdot 10^4 \cdot 50 = 2 \cdot 10^7$. In order to evaluate the number of times *P* may appear in *T* we proceed as follows. Let *P* be i.i.d. $\{x_1, \ldots, x_k\}$, and the history *T* be i.i.d. $\{y_1, \ldots, y_l\}$. Let *R* be defined as:

$$R = (l - k + 1)\binom{k}{\left\lfloor \frac{\beta}{\alpha} \right\rfloor}(2\alpha + 1)^{\left\lfloor \frac{\beta}{\alpha} \right\rfloor}\max(1, \beta \bmod \alpha) \quad (4)$$

The expectation of the number of times that the pattern *P* is included in the history is given by the following expression:

$$E(\text{number of times that } x_1, \ldots, x_k \text{ appears in } y_1, \ldots y_l) =$$
$$E\left(\sum_{i=1}^{l-k+1}[y_i, \ldots, y_{i+k-1} = x_1, \ldots, x_k]\right) = \sum_{i=1}^{l-k+1} P(y_i, \ldots, y_{i+k-1} = x_1, \ldots, x_k) \quad (5)$$

We can simplify this expression, by noticing the following:

$$\sum_{i=1}^{l-k+1} P(y_i, \ldots, y_{i+k-1} = x_1, \ldots, x_k) =$$

$$= (l - k + 1)\binom{k}{\left\lfloor \frac{\beta}{\alpha} \right\rfloor}(2\alpha + 1)^{\left\lfloor \frac{\beta}{\alpha} \right\rfloor}(\max(1, \beta \bmod \alpha)) \cdot P(y_1, \ldots y_k = x_1, \ldots, x_k) =$$

$$= R \cdot \prod_{i=1}^{k} P(y_i = x_i) = R \cdot \left(P(y_1 = x_1)\right)^k \quad (6)$$

On the other side, we have

$$R \cdot \left(P(y_1 = x_1)\right)^k = R \cdot \left(\sum_{i=1}^{10^d} P(\{y_1 = \text{some letter}\} \cap \{x_1 = \text{the same letter}\})\right)^k \quad (7)$$

Consequently, we can write that it equals the following:

$$R \cdot \left(\sum_{i=1}^{10^d} P(y_1 = \text{some letter}) \cdot P(x_1 = \text{the same letter})\right)^k = R \cdot \left(\sum_{i=1}^{10^d} \frac{1}{10^{2d}}\right)^k \quad (8)$$

In our case of above-mentioned example for length $l$=1440 (minutes in one day), we obtain that the expectation is $1436 \cdot 2 \cdot 10^7/10^{10} \approx 3$. We can even obtain a better bound for this expectation if we use inclusion-exclusion principle. In this case the number of choices with $\beta$

$$\beta \geq \sum_{i=0}^{\beta/\alpha}(-1)^i \cdot \binom{k}{i} \cdot \binom{\beta - i\alpha - 1}{k-1} \tag{9}$$

and therefore

$$R = (l - k + 1) \cdot \sum_{i=0}^{\beta/\alpha}(-1)^i \cdot \binom{k}{i} \cdot \binom{\beta - i\alpha - 1}{k-1} \tag{10}$$

For the given specific parameters provided above, we obtain that the expectation is at least 9. We should note that this evaluation takes into account only the case when the parameter $\alpha$ controls the possible difference between each measurement in $P$ versus corresponding measurement in $T$ by original value (and not by absolute value, which is in fact done in practice). This analysis shows that our algorithm will not suffer from cold-start problem in general setting.

### 4.2 Incremental evaluation of the algorithm

One of the interesting properties of our solution is that it allows to perform the incremental evaluations of the results avoiding recomputing the entire process over and over. For example, when we iterate from the current pattern $P$ to the next pattern $P'$, we know that the first element of new pattern should be deleted, and the new element (the last one) should be inserted. It means that by brute-force approach, we again need to recompute the function π for the new pattern $P'$ from the beginning. Fortunately, we can do it much faster as follows. We do not delete the first element from the new pattern, but only insert the last element (in other words, the pattern grows up by one measurement). In this case, the function π should be updated only once – for the newly inserted last element. Moreover, in the process of the search itself, during the comparison between the measurements in $T$ with those that are located in new pattern $P'$, we will make sure that the first element of the new pattern $P'$ can match any element of $T$, thus, in fact, ignoring the first element of $P'$. We can continue with this approach over and over, generating new patterns, and ignoring each time the next element in the prefix of the current pattern. Eventually, when the size of current pattern $P'$ grows up to twice size (or some constant times) of initial pattern, we can restart the next pattern having the initial size length and to compute the original function π. This allows us to significantly speed up the entire process.

## 5 Evaluation

We have evaluated our technique on DARPA 2009 dataset from IMPACT Cyber database. The DARPA 2009 dataset is created with synthesized traffic to emulate traffic between a /16 subnet (16/172.28.0.0) and the Internet. This dataset has been captured in 10 days between the 3rd and the 12th of November of the year 2009. It contains, in particular, synthetic DNS background data traffic. The dataset is large (over 6Tb) and has a variety of security events and attack types that describes the modern style of attacks. In particular it contains the events related to the DDoS and compromising DNS servers attacks. This dataset has been already evaluated using the supervised learning techniques; see [23].

The features we have looked at are:

- **Feature A**: The total number of DNS packets per minute in the traffic.
- **Feature B**: The number of malformed received DNS packets per minute, per each IP.
- **Feature C**: The number of transmitted DNS packets per minute, per each IP.

In order to evaluate the efficiency and the accuracy of our approach, we have compared it with auto-regression model (also, unsupervised method) and with other common regressors such as Linear Regression [28], Lasso [29], Random Forest Regressor [30] and KNN Regressor [31]. The common regressors have trained with the predicted window lag. We note here that the use of auto-regression as a baseline is not new, see for example [24-27]. The validation of our results has been done using standard PC server with 12 cores, 32G RAM. The entire traffic has been normalized for our string matching strategy. The normalization process has been performed according to the current logarithmic value of the average number of seen (per feature) packets. Consequently, the current error threshold $error_{threshold}$ that we have used for error evaluation was computed as a squared value of $\log_{10-\varepsilon} maxvalue$, where *maxvalue* is the current maximal number of seen (per feature) packets. Finally, parameters $\alpha$ and $\beta$ were defined as:

$$\alpha = error_{threshold} \cdot (1+\varepsilon) \cdot \frac{mean(P)}{|P|}, \beta = error_{threshold} \cdot mean(P) \quad (11)$$

where the value of $\mathrm{mean(P)}$ denotes the average number of packets in pattern *P*.

In order to evaluate the influence of each feature on the analyzed data traffic, we have associated an unique numerical value for each of the features. Feature A has a score of 1, Feature B has a score of 2 and Feature C has a score of 4. In this way, we can control the relevance of specific feature for the seek anomalies. For example, if we require the total score of currently evaluated traffic be larger than, e.g. 4, it means that we require that mean squared error and the cosine similarity will go over the thresholds for Feature C, and at least either Feature A or Feature B. The evaluated dataset has been provided with the ground truth events according to the intervals of time for their appearances. If the anomaly has been detected by our method within time interval that overlaps with the corresponding time interval in ground truth events file, we report this as true positive. If there is no such overlapping interval in ground truth events file, we report this as a false positive event. If there is any interval in ground truth events file which was not hit by any of our identified intervals of anomalies, it is considered as false negative event. All other cases are treated as true negatives.

We have also investigated how the consideration of the length of lookback history correlates with the running time of the evaluated solutions and the precision of our obtained results.

Figure 2 contains True Positive Rate (TPR), False Negative Rate (FNR), F1 (harmonic mean of precision and sensitivity) measures for all methods: Approximate String Matching, Auto-Regression, Linear Regression, Lasso, Random Forest and KNN with lookback history length (in days) from set of {0.04, 0.08, 0.25, 0.5, 0.75, 1, 2, 3, 4, 5} days and scores larger than {4,5}. ASM abbreviation stands for Approximate String Matching solution, AR stands for Auto-Regression solution, LR for Linear Regression, RF for Random Forest and KNN for K-Nearest Neighbors. The graphs for FNR, TPR and F1 score (Fig. 2) clearly support that our ASM method is quite stable in regard with the lookback history length, opposite to the others.

In the Auto-Regression model, Linear Regression and Lasso the longer history mean better results which are still worse when comparing with ASM, while the random forest behaves in

(a)

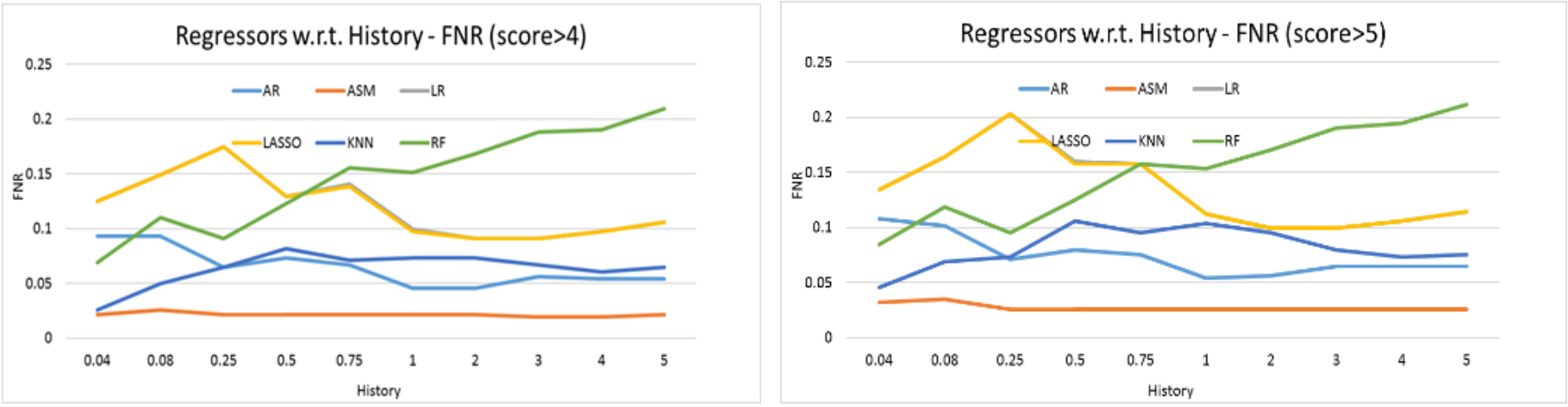

(b)

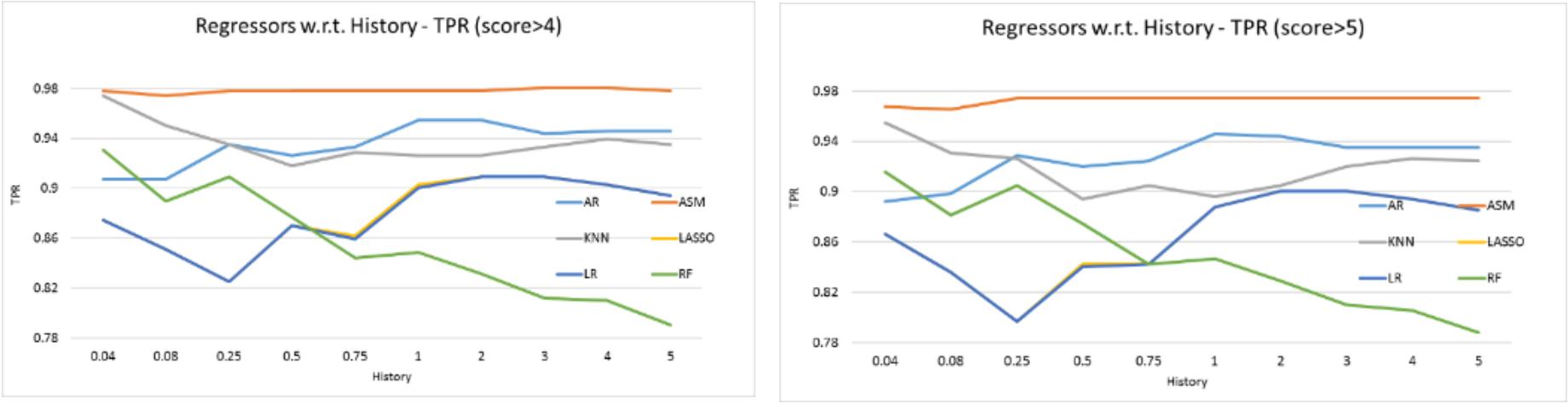

(c)

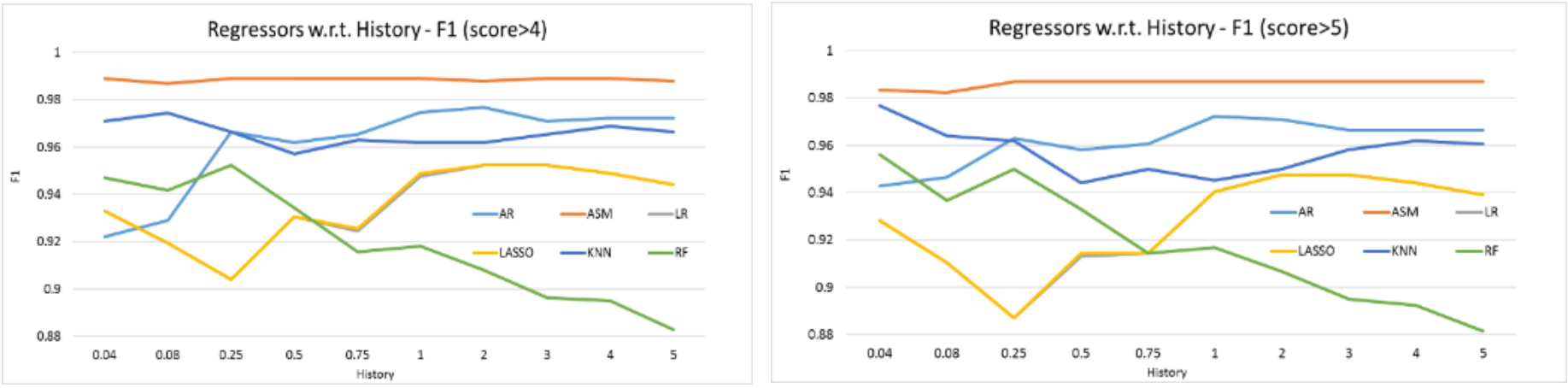

**Fig. 2.** (a) FNR (b) TPR (c) F1 for score 4,5 for all methods. ASM outperformed the other methods for all measures.

the opposite trend. The second stable method was the AR, although it performed similar to the KNN method. F1 score shows almost perfect precision and recall for ASM.

Table 1 below shows the absolute values for the mean FN, FP errors over all history for all methods. From Table 1 we can learn that in general, over all considered lookback history lengths, our method is much better than the others. The linear regression methods (LR and Lasso) performed the best false-positive (0), while their false negative were almost the worst. The AR performed worse FN but good (29.9 – 34.3) FP, while the ASM significantly reduces the number of FP and FN decisions (from up to 12 times for FP to 3 times for FN).

| | Mean FP | | Mean FN | |
|---|---|---|---|---|
| Scores > | 4 | 5 | 4 | 5 |
| AR | 5.1 | **0** | 29.9 | 34.3 |
| ASM | 0.4 | **0** | **10** | **12.7** |
| LR | **0** | **0** | 55.7 | 62.5 |
| LASSO | **0** | **0** | 55.5 | 62.4 |
| KNN | 1.5 | **0** | 29.3 | 37.8 |
| RF | 1.6 | **0** | 67.4 | 69.5 |

**Table 1.** Mean False Positive and False Negative decisions for all methods

For the Auto-regression method we evaluated the calculated lags by the AR. The results are shown in Figure 3. The lags are identical for all IPs, while the lags are larger for longer history.

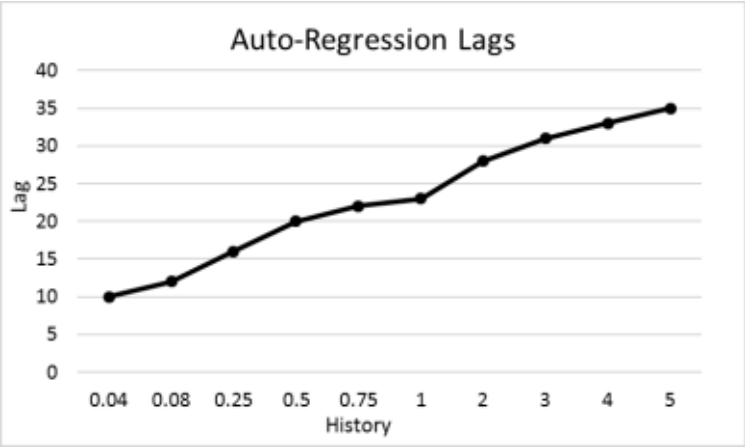

**Fig. 3.** Auto-Regression lags w.r.t history length

Additionally, we have evaluated the raw traffic for a few IPs to see the anomalies. Figure 4 shows the sum of packets for IP 172.28.10.6 which serves as the Firewall. The traffic trend is periodical, where from 14:00PM to 12:00AM there are ~200K packets per hour and from 12:00AM to 14:00PM there are ~75K packets per hour. At the 3-Nov and at the 12-Nov there are two peaks in traffic, which are reported as DNS attacks and are shown in Figure 4.

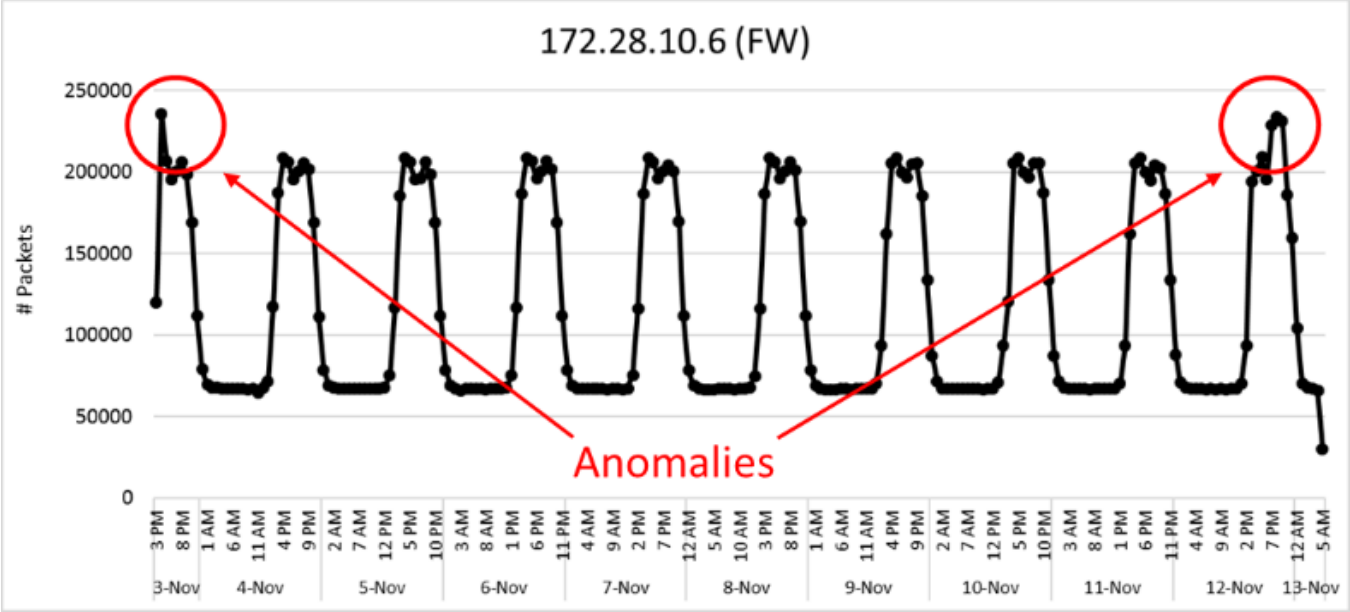

**Fig. 4.** Traffic of specific IP with 2 anomalies

Figure 5 below shows the relation between the anomaly events found by our method (ASM) and the actual anomaly events in the system (GT) for the case of lookback history of length 1. In the upper, attacks related to IP 172.28.108.88 and in the bottom DOS attacks. As we can see, there is a difference of some single events per each day.

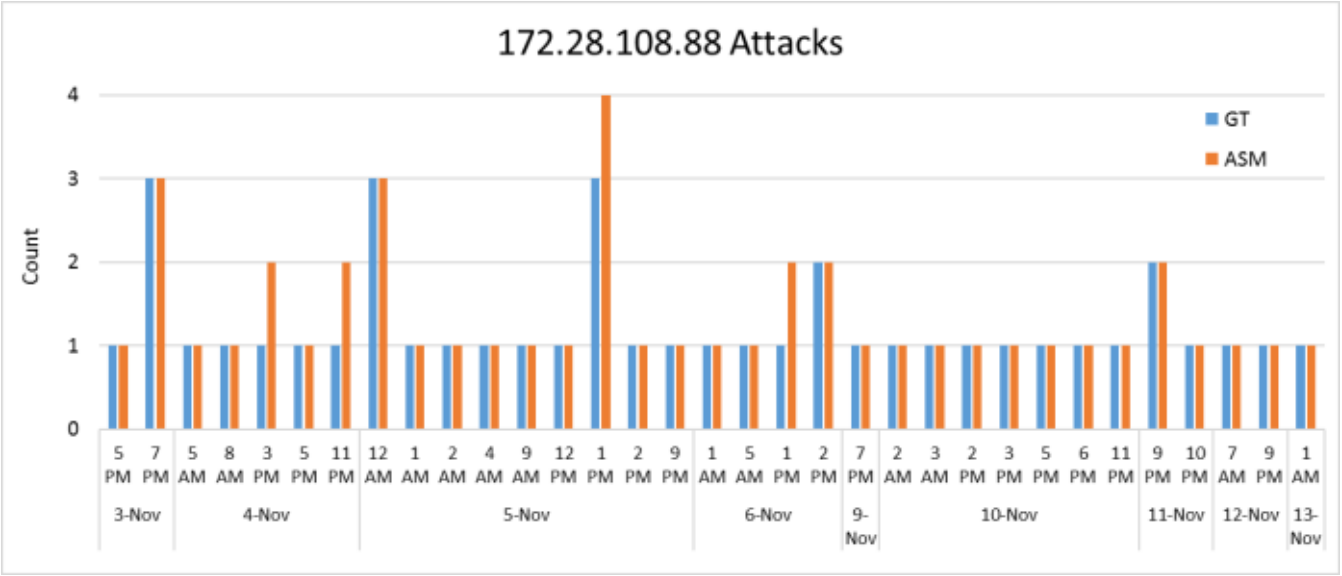

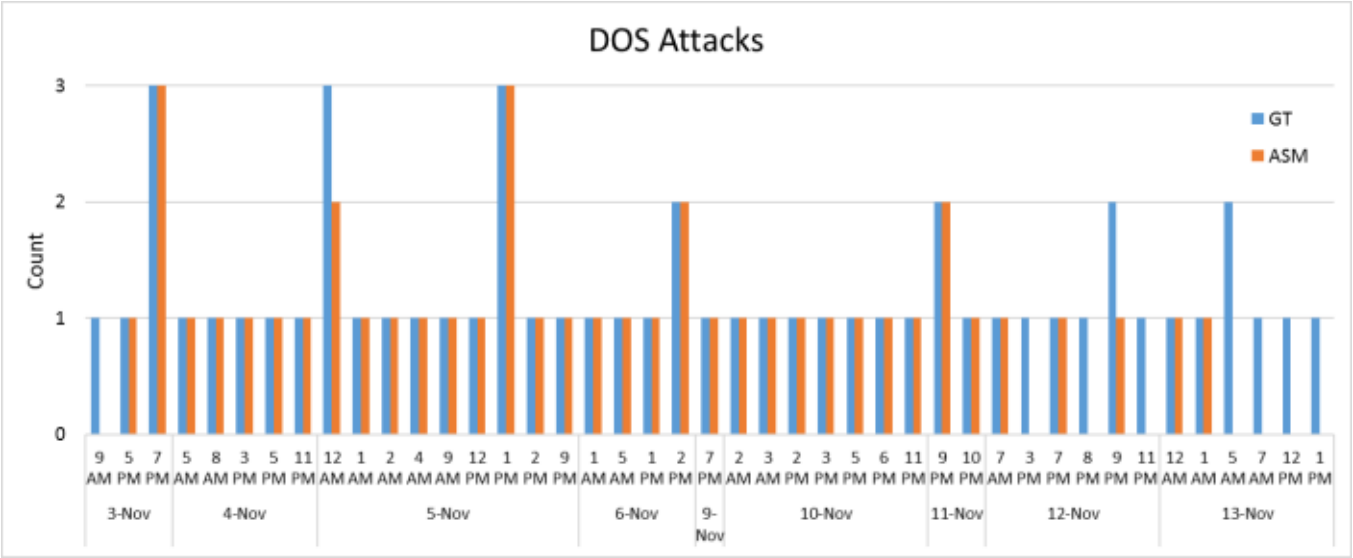

**Fig. 5.** Relation between ASM and ground truth events

We also have evaluated the processing time for all the evaluation methods. First, we calculated the processing time for the entire dataset running, as shown in Figure 6(a) below; this is useful for offline scenario, when an organization wants to find anomalies on all historical data it has. In this scenario we train a model (when it is required by the evaluation method) every 1 minute for the updated data (shift in 1 minute), and predict the future values using this model. Obviously, if a method has an incremental mode like our solution, it is much faster than the others. In Figure 6, the results without incremental mode are presented. Alternatively, we show in Figure 6(b), the 1-time query processing time which is useful for online scenario.

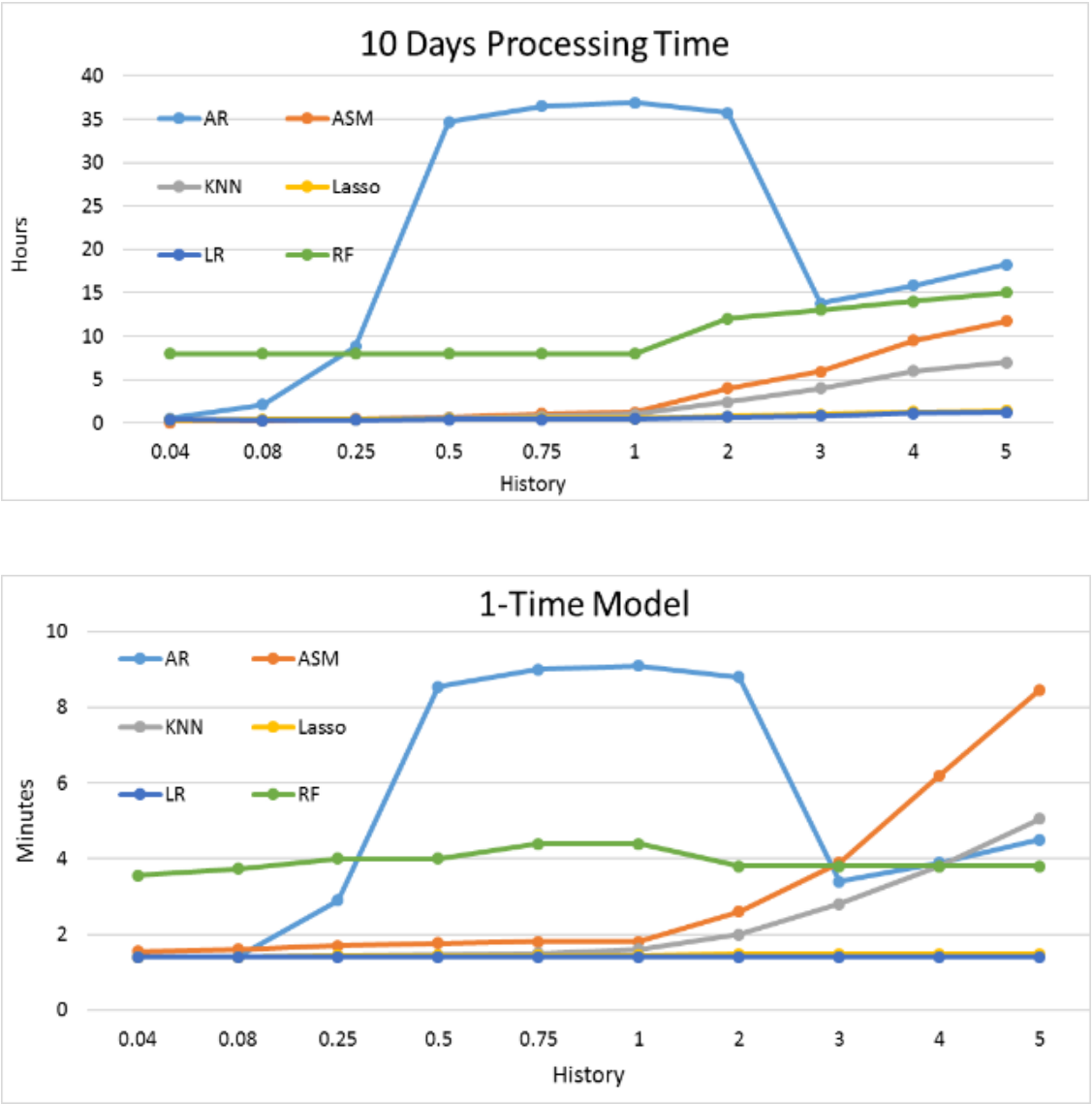

**Fig. 6.** (a) Entire dataset processing time for all methods and (b) 1-time query processing time

In the 10 days processing time, the auto-regression technique performed worst in terms of running time, and it is very inefficient in comparison to the others including our solution. For example, for half-day lookback history, auto-regression evaluation takes more than one day of run for the offline scenario, while the approximate string matching is near order of magnitude faster, taking only a few minutes. In the 1-time query processing time the results are similar, while our solution runs quite fast and can be deployed as the online solution due to its incremental mode option.

# 6 Conclusions

In this paper we have introduced and analyzed the performance of approximate string matching as one of the unsupervised machine learning techniques applied to the problem of detecting anomalies in DNS traffic. Our method is quite generic and can be applied to other types of traffic as well. Our analysis has shown a superiority of our method (both in terms of anomaly detection precision and running time) over the standard unsupervised auto-regression model as well as against other common regressors such as Linear Regression, Lasso, Random Forest and KNN. One of possible extensions could be generating the combined features normalized data (instead of looking at them separately and applying a total score mechanism). Another possibility is to consider the traffic offline data in reverse order.